\documentclass[printer]{aa}

\newcommand{\etal}{\hbox{et~al.}}
\newcommand{\msol}{\hbox{\,${\rm M_{\sun}}$}}

\newcommand{\cms}{\hbox{\,${\rm cm^{-2}}$}}

\newcommand{\cmc}{\hbox{\,${\rm cm^{-3}}$}}
\newcommand{\kms}{\hbox{\,${\rm km\,s^{-1}}$}}

\newcommand{\kcmc}{\hbox{\,${\rm K\, cm^{-3}}$}}
\newcommand{\up}{\hbox{$U$}}
\newcommand{\nh}{\hbox{$N_{H}$}}
\newcommand{\nhi}{\hbox{$N_{HI}$}}
\newcommand{\nhii}{\hbox{$N_{HII}$}}
\newcommand{\nciv}{\hbox{$N_{CIV}$}}
\newcommand{\map}{\hbox{{\sc mappings i}c}}
\newcommand{\gam}{\hbox{$\Gamma$}}
\newcommand{\gamb}{\hbox{$\Gamma$}}
\newcommand{\gamo}{\hbox{$\Gamma$}}
\newcommand{\gamc}{\hbox{$\Gamma$}}
\newcommand{\vs}{\hbox{$V_{shock}$}}
\newcommand{\zce}{\hbox{$Z_{C}^{emi}$}}
\newcommand{\zca}{\hbox{$Z_{C}^{abs}$}}
\newcommand{\zsol}{\hbox{$Z_{\sun}$}}

\newcommand{\mgiiw}{\hbox{Mg\,{\sc ii}\,$\lambda\lambda $2798}}

\newcommand{\ciii}{\hbox{C\,{\sc iii}]}}
\newcommand{\ciiiw}{\hbox{C\,{\sc iii}]$\lambda $1909}}
\newcommand{\civ}{\hbox{C\,{\sc iv}}}
\newcommand{\civw}{\hbox{C\,{\sc iv}\,$\lambda\lambda $1549}}
\newcommand{\civww}{\hbox{C\,{\sc iv}\,$\lambda\lambda $1548, 1551}}
\newcommand{\nv}{\hbox{N\,{\sc v}}}
\newcommand{\nvw}{\hbox{N\,{\sc v}$\lambda $1240}}

\newcommand{\oviw}{\hbox{O\,{\sc vi}$\lambda $1035}}

\newcommand{\oiii}{\hbox{[O\,{\sc iii}]}}

\newcommand{\oiiiuvw}{\hbox{O\,{\sc iii}]$\lambda $1663}}
\newcommand{\oiiiuv}{\hbox{O\,{\sc iii}]}}

\newcommand{\oii}{\hbox{[O\,{\sc ii}]}}

\newcommand{\lya}{\hbox{Ly$\alpha$}}

\newcommand{\ha}{\hbox{H$\alpha$}}

\newcommand{\hi}{\hbox{H\,{\sc i}}}
\newcommand{\hii}{\hbox{H\,{\sc ii}}}

\newcommand{\heii}{\hbox{He\,{\sc ii}}}

\newcommand{\heiiuw}{\hbox{He\,{\sc ii}\,$\lambda $1640}}
\newcommand{\hei}{\hbox{He\,{\sc i}}}

\usepackage{graphics}

\begin{document}

\thesaurus{11(11.09.1;12.05.1;11.01.2;11.06.1;11.09.4;02.12.1) }

\title{A vestige low metallicity gas shell surrounding the radio
galaxy 0943--242 at $z=2.92$}


\author{L. Binette\inst{1}, J. D. Kurk\inst{2}, 
M. Villar-Mart\'\i n\inst{3}, H. J. A. R\"ottgering\inst{2}
          }

\authorrunning{Binette et~al.}

\offprints{Luc Binette}

\institute{Instituto de Astronom\'\i a, UNAM, 
	Ap. 70-264, 04510 M\'exico, DF, M\'exico ~(e-mail: binette@astroscu.unam.mx)  
\and  Leiden Observatory, P. O. Box 9513, 2300 RA, Leiden, The Netherlands
\and Department of Physical Sicences, University of Hertfordshire,
College Lane, Hatfield Herts, AL10 9AB, England 
}
\date{Received / Accepted}

\maketitle

\begin{abstract}

Observations are presented showing the doublet \civww\ absorption
lines superimposed on the \civ\ emission in the radio galaxy
0943--242. Within the errors, the redshift of the absorption system
that has a column density of $\nciv = 10^{14.5 \pm 0.1} \,\cms$
coincides with that of the deep \lya\ absorption trough observed by
R\"ottgering et~al. (1995). The gas seen in absorption has a resolved
spatial extent of at least 13\,kpc (the size of the extended emission
line region).  We first model the absorption and emission gas as
co-spatial components with the same metallicity and degree of
excitation. Using the information provided by the emission and
absorption line ratios of \civ\ and \lya, we find that the observed
quantities are incompatible with photoionization or collisional
ionization of cloudlets with uniform properties.  We therefore reject
the possibility that the absorption and emission phases are co-spatial
and favour the explanation that the absorption gas has low metallicity
and is located further away from the host galaxy (than the emission
line gas). The larger size considered for the outer halo makes
plausible the proposed metallicity drop relative to the inner emission
gas. In absence of confining pressure comparable to that of the
emission gas, the outer halo of 0943--242 is considered to have a very
low density allowing the metagalactic ionizing radiation to keep it
higly ionized. In other radio galaxies where the jet has pressurized
the outer halo, the same gas would be seen in emission (since the
emissivity scales as $n_H^2$) and not in absorption as a result of the 
lower filling factor of the denser condensations. This would
explain the anticorrelation found by Ojik et~al. (1997) between \lya\
emission sizes (or radio jet sizes) and the observation (or not) of
\hi\ in absorption. The estimated low metallicity for the absorption
gas in 0943--242 ($Z \sim 0.01 \zsol$) and its proposed location
--outer halo outside the radio cocoon-- suggest that its existence
preceeds the observed AGN phase and is a vestige of the initial
starburst at the onset of formation of the parent galaxy.

\end{abstract}

\section{Introduction}

Very high redshift ($z>2$) radio galaxies (hereafter HZRG) show
emission lines of varying degree of excitation. In virtually all
objects, the \lya\ line is the strongest and is usually accompanied by
high excitation lines of \civw, \ciiiw, \heiiuw\ and, at times, \nvw\
(R\"ottgering et~al. 1997 and references therein).  An important
characteristic of the emision gas is its spatial scale.  The sizes of
the \lya\ emission region range from $15$ to 120\,kpc (van~Ojik \etal\
1997).

Most ground work on HZRG is performed at rather low resolution ($\sim
20$\AA) to maximize the probability of line detection and the S/N.
However a very potent discovery was made by van~Ojik \etal\ (1997,
hereafter vO97) at much higher resolution, that of extended \hi\ {\it
absorption} gas. In effect, out of 18 HZRG spectra taken at the
unusually high resolution of $\simeq 1.5$--3\AA, vO97 found --in 60\% of
the objects-- deep absorption troughs superimposed on the Ly$\alpha$
emission profiles.  Furthermore, out of the 10 radio galaxies smaller
than 50\,kpc, strong \hi\ absorption is found in 9 of them. The
absorption gas appears to have a covering factor near unity over very
large scales, namely as large as the underlying emission gas.

The current paper addresses the problem of the ionization state of
both the absorption and the emission gas as well as the
interconnection between the two. The main justifications behind this
work are the following: HZRG are probably the progenitors of the
massive central cluster galaxies (Pentericci et~al. 1999) and as such
are an important means by which we can study large ellipticals and
their environment at such high redshift, a time not so long after, or
even during their formation. Furthermore, the extended gas as detected
in \civ\ (see below) is chemically enriched and therefore represents
debris of past intense stellar formation periods and is interesting to
study in their own right. What is the fate of such gas?  How quickly
has the enrichment of this large scale gas proceeded? Will this gas be
heated up into a hot wind and enrich the intergalactic X-ray gas in
cluster of galaxies? Will it on the contrary condense into sheets or
condensations? A better understanding of the various gaseous phases
which co-exist in high redshift objects would help anwering these
questions.

To determine the physical conditions of the absorption gas, new
observations were carried out at the wavelength of \civ\ and \heii\ in
0943--242, the first radio galaxy reported to show large scale absorption
troughs (R\"ottgering et~al. 1995, hereafter RO95).  The new spectrum
 shows the \civ\ absorption doublet at the same redshift\footnote{We will
distinguish between absorption and emission redshifts using
subscripts, as in $z_a$ and $z_e$, respectively.} $z_{a}$ as the \lya\
absorption trough (RO95). Clearly and surprisingly the gas in absorption is
highly ionized and probably of comparable excitation to the gas seen
in emission.

This paper is structured as follows. We first present observations
which show \civ\ in absorption in 0943--242 (Sect.\,~\ref{observ}). In
Sect.\,~\ref{hypo} we derive a ratio (\gamo) relating the observed
emission and absorption quantities which depends somewhat on the
ionization fraction of H but not explicitely on the C/H metallicity
ratio.  At first, we postulate that the emission and absorption gas
components are co-spatial and share the same excitation mechanism and
physical conditions and proceed to model \gamo\ with a one-zone
equilibrium photoionization model. We improve on the model using a
stratified photoionized slab. As the observed ratio cannot be
reproduced even in the case of collisional ionization, we discuss in
Sect.~\ref{intergam} two alternative interpretations of this
significant discrepancy. We demonstrate the many advantages of the
winning scenario in which the absorption gas is further out and of
much lower density, pressure and metallicity than the emission gas.


\section{Observations of \civ\ (and \lya) in absorption in 0943--242} \label{observ}

\begin{figure}
\resizebox{\hsize}{!}{\includegraphics{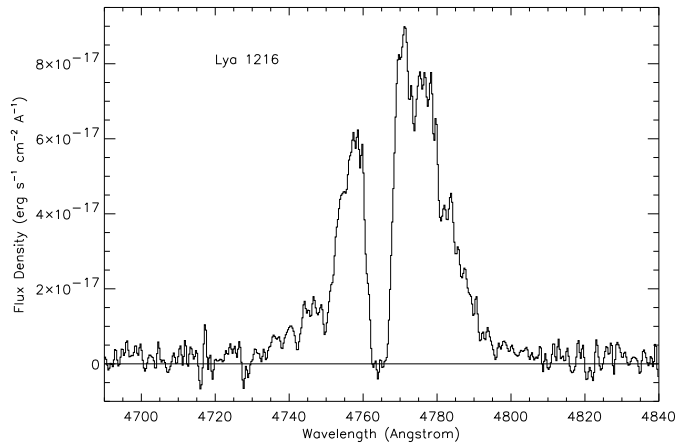}}
\caption{An expanded plot of the \lya\ spectral region obtained by
RO95.  The \hii\ emission gas redshift is $z_e=2.9233 \pm 0.0003$
and the main absorber of column $\nhi = 10^{19.0\pm 0.2}$\,\cms\ lies
at $z_a=2.9200\pm 0.0002$.}
\label{fig:lya}
\end{figure}

\subsection{Earlier observations of 0943--242 at $z_e = 2.92$ } \label{oldobs}

The low resolution spectrum of 0943--242 shown in RO95 and 
discussed also in van~Ojik et~al. (1996) displays the characteristic
emission lines of a distant radio galaxy: strong \lya, weaker \civ,
\heii\ and possibly \ciii.   This object was also observed at intermediate 
resolution (1.5\AA) by RO95 in the region of \lya\ with the slit
positioned along the radio axis. The initial discovery of extended
absorption troughs was based on this latter spectrum which we
reproduce in Fig.~\ref{fig:lya}.

\subsection{New observations of $C\,{IV}$ and  $He\,{II}$ at intermediate resolution}

With the objective of providing constraints on the abundances and
kinematics of the gas in 0943--242, sensitive high-resolution spectroscopic
observations centered at the \civ\ and \heii\ lines were   performed
at the Anglo Australian Telescope (AAT) on 1995 March 31 and April 1
under photometric conditions and with a seeing which varied from 1\arcsec\
to 2\arcsec. The RGO spectrograph was used with a 1200 grooves
mm$^{-1}$ grating and a Tektronix 1024$^2$ thinned CCD, yielding
projected pixel sizes of $0.79\arcsec\ \times 0.6$\AA. The projected
slit width was 1.3\arcsec, resulting in a resolution as measured from
the copper-argon calibration spectrum of 1.5\AA\ FWHM; the slit was
oriented at a position angle of 74$^\circ$, i.e. along the radio axis
(as in RO95).

\begin{figure}
\resizebox{\hsize}{!}{\includegraphics{luc1000.f1}}
\caption{The full-resolution AAT spectrum  showing the \civww\ and
\heiiuw\ lines.}
\label{fig:full}
\end{figure}

The total integration time of 25000s was split into 2$\times$2000s and
7$\times$3000s exposures in order to facilitate removal of cosmic
rays.  Exposure times were chosen to ensure that the background was
dominated by shot noise from the sky rather than CCD readout noise.
Between observations the telescope was moved, shifting the object slit
by about 3 spatial pixels, so that for each exposure the spectrum was
recorded on a different region of the detector. The individual spectra
were flat-fielded and sky-subtracted in a standard way using the
long-slit package in the NOAO reduction system IRAF.  The precise
offsets along the slit were determined using the position of the peak
of the spatial profile of the \civ\ and \heii\ lines. Using these offsets, the
images were registered using linear interpolation and summed to obtain
the two-dimensional spectrum. The resultant seeing in the final
two-dimensional spectrum, measured from two stars on the slit, was
1.5\arcsec\ FWHM. The corresponding FWHM of \civ\ emission along the
slit was 2.2\arcsec, giving a deconvolved (Gaussian) width of
1.6\arcsec\ or 12\,kpc. Within the errors, this is the same as that
found for \lya\ emission by RO95.

The two-dimensional spectrum was weighted summed over a 7 pixel
(5\arcsec) aperture to obtain a one-dimensional spectrum. In
Fig.~\ref{fig:full} we show the AAT data in the form of a
full-resolution spectrum.

\subsection{Profile fitting of the emission and absorption
\lya\ and $C\,{IV}$ lines}

One deep trough is observed in the \lya\ emission line
(Fig.~\ref{fig:lya}) which was interpreted as a large scale \hi\
absorber by RO95.  In addition there are a number of weaker
troughs, presumably due to weak \hi\ absorption.  Fitting the emission
line by a Gaussian and the \hi\ absorption by Voigt profiles, RO95
infer a column density \nhi\ of $ 10^{19.0\pm 0.2}$\,\cms\ for the deep
trough, a redshift $z_a=2.9200\pm 0.0002$ and a Doppler parameter $b$ of
$55\pm5$\,\kms. For the three shallow troughs, they find \nhi\ ranging
from $10^{13.8}$ to $10^{14.1}$ \cms\ and $b$ ranging from 7 to 100
\kms. The redshift difference of the absorbers relative to systemic
velocity when converted into inflow/outflow velocities indicate values
not exceeding 800\,\kms.  Because at the bottom of the main trough no
emission is observed, the covering factor of the absorbing gas must be
equal or larger than unity over the complete area subtended by the
\lya\ emission, indicating that the spatial scale of the absorber
exceeds 13\,kpc. This work will concern only the deep absorption
trough.

\begin{figure}
\resizebox{\hsize}{!}{\includegraphics{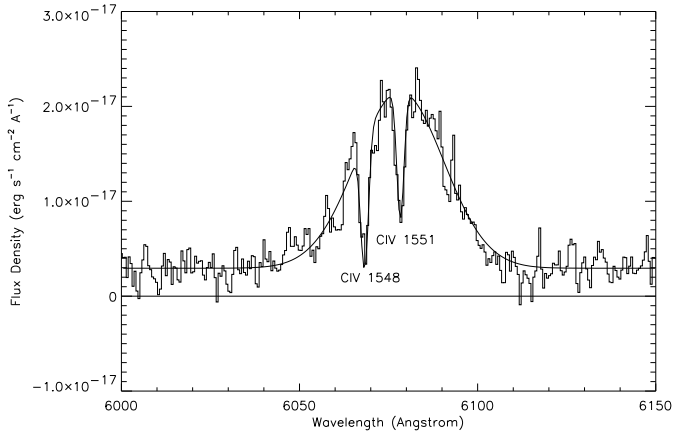}}
\caption{An expanded plot of the full-resolution spectrum 
with the fit superimposed (solid line).}
\label{fig:CIVfit}
\end{figure}

To parameterize the \civ\ profile we have assumed that the underlying
emission line is Gaussian, with Voigt profiles due to the
\civ\ doublet absorption superimposed. We used an iterative
scheme that minimizes the sum of the squares of the difference between
the model and the observed spectrum, thereby solving for the
parameters of the model (e.g.\ Webb 1987, vO97). Initial values were assumed
for the shape of the Gaussian profile and the redshift of the
absorber.

\begin{table}
\caption{\label{ta:parameters}Parameters for the Gaussian and Voigt
profile fits}
\label{ewciv}
\begin{flushleft}
\begin{tabular}{l@{}rr} \hline
{\bf Emission} & {\bf \civ} & {\bf \heii} \\ \hline
Offset ($10^{-17}$ erg cm$^{-2}$ s$^{-1}$) &
0.29   $\pm$ 0.01 & 0.32  $\pm$ 0.07 \\
Peak ($10^{-17}$ erg cm$^{-2}$ s$^{-1}$) &
1.90   $\pm$ 0.1  & 1.75  $\pm$ 0.2 \\
Position of peak (\AA) &
6078.2 $\pm$ 0.5 & 6434.5 $\pm$ 0.5 \\
$z_e$ & 2.9247 $\pm$ 0.0003 & 2.925 $\pm$ 0.001 \\
FWHM (\kms) &
1430     $\pm$ 50   & 1025     $\pm$ 45 \\ \hline
{\bf Absorption} & {\bf \civ} & \\ \hline
$z_a$ & 2.9202 $\pm$ .0002 & \\
$b$ (\kms) &  45      $\pm$ 15 & \\
\nciv\ (cm $^{-2}$)&  10$^{14.5 \pm 0.1}$ & \\
Position of $1^{\rm st}$ trough (\AA)& 6068.2 & \\
Position of $2^{\rm nd}$ trough (\AA)& 6078.3 & \\ \hline
\end{tabular}
\end{flushleft}
\end{table}

In Fig.\,\ref{fig:CIVfit} we show a portion of the spectrum with the
model fits superimposed. The Gaussian fitted to the \civ\ emission
line peaks at $z_e = 2.9247 \pm 0.0003$ and has a  FWHM of
$29\pm2$\AA. We have corrected all wavelengths to the vacuum
heliocentric system ($\simeq $+1.13\,\AA) before computing the redshifts.
The two troughs in this figure correspond to the
\civww\  doublet produced by the same absorption
system. Therefore, within the fitting procedure, the wavelength
separation and the ratio of the two profiles' depths are fixed by
atomic physics while the two values for $b$ are set to be equal. The
fit gives for the location of the bottoms of the two troughs
$\lambda=$ 6068.2 and 6078.3\AA\ resulting in a redshift of 2.9202
$\pm$ 0.0002. Within the errors this redshift is equivalent to that of
the main \hi\ absorber and in the subsequent analysis we will assume
that the \lya\ and \civ\ {\it absorption} gas belongs to the same
absorber. We derive a Doppler parameter $b$ for the doublet of $45
\pm 15 \kms$ and a column density \nciv\ of 10$^{14.5 \pm 0.1}\,\cms$
as summarized in Table~\ref{ewciv}.

As expected, \heii\ appears only in emission without any
absorption since it is not a resonance line. Parameters for the
\heii\ emission profile were obtained by fitting a Gaussian
using the same iterative scheme (see Fig.~1 in R\"ottgering \& Miley
1997). The peak is positioned at $z_e = 2.925 \pm 0.001$ and has a FWHM
of $22 \pm2$\AA. The fitted parameters of the emission and
absorption profiles are presented in Table~\ref{ewciv}. We recall that
the FWHM of the \lya\ emission profile  is $1575 \pm 75 \,
\kms$ (vO97), significantly larger than that of \heii\ 
(see Table~\ref{ewciv}). Inspection of the various profiles in
Fig.~\ref{fig:lya} and Fig.~\ref{fig:full} (or Fig.~\ref{fig:CIVfit})
suggests the presence of an excess flux on the blue wings of all the
emission profiles.  Combining information from all the emission lines,
our best estimate of the emission gas redshift is $z_e=2.924\pm
0.002$.

\subsection{Velocity shear and subcomponents} \label{shear}

\begin{figure}
\resizebox{\hsize}{!}{\includegraphics{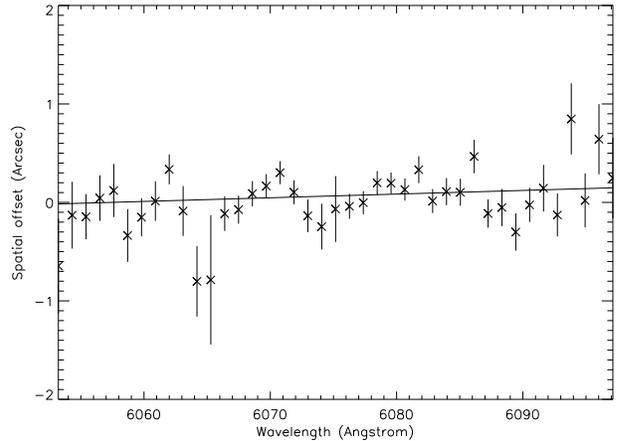}}
\caption{The relative location of the peak of the \civ\ 
emission at constant wavelength as a function of wavelength. The line
is a weighted fit to these peaks. The zero offset is arbitrary.}
\label{fig:relloc}
\end{figure}

To investigate whether there is any velocity shear in the \civ\
emission profile we fitted spatial Gaussian profiles to the emission
line as a function of wavelength. In Fig.\ \ref{fig:relloc} we show
the wavelength maxima of these spatial profiles and a line fitted
through these points. The spatial profile of the \civ\ emission
spectrum is displaced by 0.2\arcsec, corresponding to a displacement
of 1.5 kpc, over a wavelength range of 50\AA. RO95 measured a
comparable shift for \lya\ of 1.8 kpc\footnote{This new value of $0.33
\pm 0.06 \, {\rm pixels} \times 0.74 \, {\rm arcsec/pixel} = 
0.2442 \, {\rm arcsec} \times 7.36 \, {\rm kpc/arcsec} = 1.80 \pm 0.33$\,kpc
is to be preferred to that quoted by RO95 of 2.5\,kpc.}  although it
appears that the latter displacement is due to a far more pronounced
and abrupt difference in locations of the \lya\ peak on both sides of
the absorption trough. As Fig.\,\ref{fig:relloc} shows, the peaks of
\civ\ emission form a wavy line. We believe the velocity shear in the
C{\sc iv} profile to be less significant than the shear in the
\lya\ profile. We cannot rule out that the small velocity shear might
be masking a possible break up of the absorption regions into a few
saturated absorption components of smaller $b$.

%

A concern about the determination of \nciv\ is the possibility that
that there exist subcomponents in the absorption systems that have
high column densities but low $b$ values and are, therefore, not
acounted for whenever individual velocity subcomponents are not
resolved. Although we cannot strictly exclude this possibility, we
adopt the stand of Jenkins (1986) and Steidel (1990a) who, using
extensive absorption line studies, argue that this is unlikely to be
the case, at least for \civ, and that a single-component
curve-of-growth analysis can be used to infer total columns
although the inferred {\it effective} $b$ value has no physival meaning in
terms of temperature. It is interesting to note that the
physical conditions inferred from the \civ\ fit are fully consistent with
the observed ratio of the doublet (since both troughs are equally well
fitted). If the underlying continuum was flat, the \nciv\ column and
the $b$ value we infer would imply a theoretical ratio of
equivalent widths of $W_0(1548)/W_0(1551) = 1.4$, which is where the
curve of growth just begins to leave the linear part (Steidel 1990a).
Clearly the \nhi\ column might be susceptible to a larger  error since
\lya\ is saturated. With these caveats in mind, we will assume in the
following analysis that the adopted columns do not lie far off from
reality.

\section{A simple model for the ionized gas in emission and absorption} \label{hypo}

Our initial hypothesis is that the absorption gas is a subcomponent of
the emission gas, sharing the same excitation mechanism and
metallicity. We discuss the physical conditions of such gas and
proceed to calculate an observable quantity, \gam, against which to
compare the information provided by the \lya\ and \civ\ lines in
0943--242.

\subsection{Relation between the ionized absorption 
and emission components}

The \civ\ and \lya\ lines are both resonant lines and therefore prone
to be seen in absorption against a strong underlying source. This
property has consequences for the emission gas as well. In
effect, for a geometry consisting of many condensations for which the
cumulative covering factor approaches unity, the resonant line photons
must scatter many times in between the condensations before they can
escape. In this case, the emerging flux of any resonant line from a
non uniform distribution of gas will not in general be an isotropic
quantity but will depend on geometrical factors and on the relative
orientation of the observer, a point which we now develop further.

We propose that some kind of asymmetry within the emission gas
distribution can explain how a fraction of the ionized gas can be
seen in absorption against other nearby components in emission. Let us
suppose that the emission region is composed of low filling factor
ionized gas condensations which are denser (therefore brighter)
towards the nuclear ionizing source. In this picture, the \lya\ or
\civ\ photons are generated within and escape from such condensations,
after which they start scattering on the surface of neighboring
condensations until final escape from the galaxy (we assume that the
cumulative covering factor is unity). Let us now suppose an
asymmetry\footnote{The asymmetry would take place either in space or
in velocity domain or both.} in the global distribution of the outer
condensations respective to the plane of the sky. In this case, the
total number of scatterings on neighboring condensations before final
escape will differ depending on the perspective of the absorber.
Since for an observer situated on the side with an excess of
condensations many of the resonant photons would have been
`reflected' away, we expect that the reduced flux would appear as an
absorption line at the same velocity as that of the condensations
responsible for reflecting away the resonant photons.  The outer
condensations (responsible for the absorption) must necessarily be of lower
density in order to be of negligible emissivity respective
to the inner (denser and therefore brighter) emission gas, otherwise
the outer gas would out-shine in emission!

We should point out that for a density of the absorption gas as high
as 100\cmc\ as argued for in vO97, such a gas cannot be photoionized
by the metagalactic background radiation which would be much too
feeble to produce \civ. The ionization to such a degree of the
absorption gas is in itself puzzling. We adopt as working hypothesis
that it is --similarly to the emission gas-- photoionized by the AGN
or by the hard radiation from photoionizing shocks.

Finally, the fact that both the absorption and emission gas contain a
significant amount of C$^{+3}$ argues in favor of a common geometry
and excitation mechanism for the gas, the underlying hypothesis behind
the calculations developed below.


\subsection{The observable quantity \gam\ }

The quantities determined from observation of 0943--242 are the
following: the emission line ratio measured by R\"ottgering
et~al. (1997) is ${I_{CIV}\over I_{Ly\alpha}} = 0.194$. We adopt the
value of 0.17 following estimation of the missing flux due to the
absorption troughs. As for the absorption gas, the \hi\ and \civ\
column densities are $10^{19}\,\cms$ and $10^{14.5}\, \cms$, respectively, as
discussed in Sect.~\ref{observ}.  These four quantities carry
information on the three ionization species H$^{0}$, H$^{+}$ and C$^{+3}$. We
define the ratio \gamo\ as the following product of the emission and
absorption ratios:

\begin{equation}
\gamo = {I_{CIV}\over I_{Ly\alpha}} {N_{HI} \over N_{CIV}} = 0.17
{10^{19.0}\over10^{14.5}} \simeq 5400
\end{equation}

\noindent where $\nhi/\nciv$ is the ratio of the
measured absorption columns.  If, as postulated above, the gas
responsible for absorption is simply a subset of the line emitting
gas, the ratio \gam\ does not explicitly depend on the abundance of
carbon as shown below.

\subsection{The simplest case of an homogeneous one-zone slab}

To compute \gamc, in a first stage let us consider an homogeneous slab
of thickness $L$ of uniform gas density, temperature and ionization
state to represent both the gas in emission and in
absorption. Ignoring any peculiar scattering effects, the emission
line ratio ${I_{CIV}\over I_{Ly\alpha}} $ is given by the ratio of the
local emissivities $j_{CIV}/j_{Ly\alpha}$ since the slab is
homogeneous. For the emissivity of the \civ\ line, we have

\begin{eqnarray}
4 \pi j_{CIV} = 8.63\, 10^{-6} \, h\nu_{C_{\sc iv}} \, n_e n_{CIV}
 \nonumber \\ \times {\Omega_{C_{\sc IV}} \over \omega_1} 
\exp{(-h\nu_{CIV}/kT)}/\sqrt{T}
\end{eqnarray}

\noindent (Osterbrock 1989) 
where $T$ is the temperature, $\Omega_{C_{\sc IV}}$ the collision
strength of the combined doublet, $ \omega_1$ the statistical weight
of the ground state and $h\nu_{CIV}$ the mean energy of the \civ\
excited level.  For the \lya\ emissivity, we have

\begin{equation}
4 \pi j_{Ly\alpha} =  h\nu_{Ly\alpha }  \, n_e n_{HII} \, \alpha_{2p}^{eff}(T) 
\end{equation}

\noindent where $\alpha_{2p}^{eff}$ is the effective recombination
coefficient rate to level $2p$ of H (Osterbrock 1989). By putting the
temperature dependence and all the atomic constants in the function
$f(T)$, the emission line ratio becomes:

\begin{equation}
{I_{CIV}\over I_{Ly\alpha}} = {  \zce n_H \eta_{CIV} \over  n_H y_{HII}} f(T) \label{eqemi}
\end{equation}

\noindent where $n_H$ is the total hydrogen density, \zce\ the carbon
abundance relative to H of the emission gas, $\eta_{CIV}$ the
fraction of triply ionized C and $y_{HII}$ the ionization fraction of
H. 

The ratio of column densities \nhi/\nciv\ can be written as:

\begin{equation}
 {N_{HI} \over N_{CIV}} = {  n_H
x_{HI}  \over  \zca n_H \eta_{CIV} } \label{eqabs}
\end{equation}

\noindent where $x_{HI}$ is the neutral fraction of H inside our
homogeneous slab and \zca\ the carbon abundance of the {\em absorption}
gas. As we are testing the case which equates the absorption gas with the
emission gas,  then $\zca=\zce$. We denote as \gamc\ the product of
the two calculated ratios:

\begin{equation}
\gamc = {I_{CIV}\over I_{Ly\alpha}} {N_{HI} \over N_{CIV}} =
{ x_{HI} \over y_{HII} } f(T) \label{eqstd}
\end{equation}

\noindent We note that \gamc\ is not directly dependent on either 
the abundance of C or on its ionization state. It is, however,
dependent on the temperature and on the ionization state of H through
the ratio\footnote{For all practical purposes, the high ionization
regime under consideration implies that $y_{HII} = 1$.} ${ x_{HI}
\over y_{HII} }$.  To compute
this ratio, it is necessary to postulate an excitation mechanism.  For
this purpose, we have used the code \map\ (Binette, Dopita \& Tuohy
1985; Ferruit et~al 1997) to compute ${ x_{HI} \over y_{HII} }$ under
the assumption of either collisional ionization or
photoionization. Here are the results.

\begin{enumerate}

\item{\it Photoionization.}
Putting in the atomic constants and calculating the equilibrium
temperature and ${ x_{HI} \over y_{HII} }$ in the case of
photoionization by a power law of index $\alpha$ ($F_{\nu} \propto
\nu^{\alpha}$) of either $-0.5$ or $-1$, we find that the calculated 
\gamc\ always lies within the range 0.8--12. The explored range in
ionization parameter\footnote{We use the customary definition of the
ionization parameter $\up\ = \varphi_H/n_H$ as the ratio between the
density of ionizing photons (impinging on the slab) and the total H
density at the face of the slab.} \up\ covered all the values which
produce significant \civ\ in emission (${\civ}/C > 8$\%), that is
$10^{-3.5} < \up\ < 10^{-1}$.

\item{\it Collisional ionization.}
In this sequence of models, we calculated the ionization equilibrium
of a plasma whose temperature varied from 30\,000\,K to 50\,000\,K.
We find that \gamc\ remains in the similar low range of 6--13. At the
lower temperature end, \lya\ emission is enhanced considerably by 
collisional excitation, which contributes in reducing \gamc.

\item{\it Additional heating sources.} To cover the case of
photoionization at a higher temperature than the equilibrium value
(due to additional heating sources such as shocks), we artificially
increased the photoionized plasma temperature to 40\,000\,K or
50\,000\,K for calculations with the same values of \up\ as
above. This did not extend the range of \gamc\ obtained.

\end{enumerate}

We conclude that for the simple one-zone case, \gamc\ consistently remains
below the observed value by more than two orders of magnitude.

\begin{figure}    
\resizebox{\hsize}{!}{\includegraphics{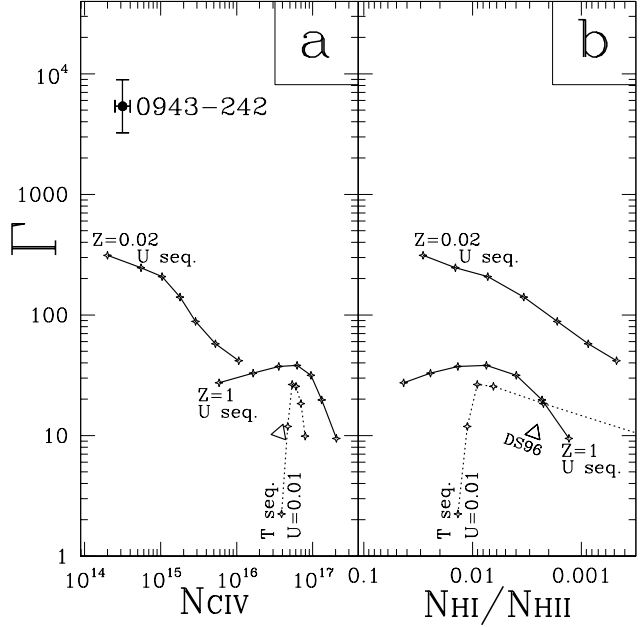}}
\caption[]{{\bf a:} Calculated and observed \gamb\  as a function
of the column density \nciv. The filled circle represents the observed
value for 0943--242.  {\bf b:} The same models as a function of the
column ratio \nhi/\nhii. In both panels, the solid line represents a
sequence of photoionized slabs with \up\ increasing from left to
right, starting at $10^{-2.5}$. The gas total metallicity is either
solar ($Z=1$) or 1/50th solar. The separation between tick marks
corresponds to an increment of 0.25\,dex in \up.  All slab
calculations were truncated at a depth corresponding to the observed
$\nhi = 10^{19}\,\cms$. The slab total column or $\nhii$ can be
inferred from panel~b. [If we were to reduce by 100 the abundance of
the absorption gas while keeping solar the emission gas ($\zca/\zce =
0.01$, see Eqs~\ref{eqemi} and \ref{eqabs}), this would be equivalent
to translating by 2\,dex both up and to the left the $Z=1$ sequence of
panel~a.] The dotted line represents a sequence of slabs of arbitrary
uniform temperatures (all with $\up = 10^{-2}$ and $Z=1$) covering the
range 10\,000\,K to 40\,000\,K (from left to right) by increments of
0.1\,dex in $T$.  The open triangle represents a slab photoionized by
a high velocity shock of \vs\ = 500\,\kms\ from Dopita \& Sutherland
1996. }
\label{omeg}
\end{figure}

\subsection{The ionization stratified slab} \label{upseq}

To verify whether a stratified slab geometry might alter the above
discrepancy in \gam, we have calculated in a similar fashion to
Bergeron \& Stasi\'nska (1986) and Steidel (1990b) the internal
ionization and temperature structure of a slab photoionized by
radiation impinging on one-side (i.e. one-dimensional ``outward only''
radiation transfer) using the code \map. We adopted a power law of
index $\alpha =-1$ as energy distribution. Since the column densities
of H and C are useful diagnostics on their own right, we present in
Fig.~\ref{omeg} the value of \gamb\ for a slab as a function of \nciv\
(left panel) and \nhi/\nhii\ (right panel). (One can interpret
\nhi/\nhii\ of Panel~b as the mean neutral fraction of the slab:
$\left<{x_{HI} / y_{HII}}\right>$.)

The solid line in Fig.~\ref{omeg} represents a sequence of different
slab models with increasing ionization parameter from left to right
covering the range $10^{-2.5} \le \up \le 10^{-1}$ for a gas of either
solar metallicity ($Z=1$) or with a significantly reduced metallicity of
${1}/{50}$th solar.  The practical constraint that \civ\ be a strong
emission line implies that $\up \ge 10^{-2.5}$. In all calculations,
the thickness of the slab is set by the observable condition that
$\nhi = 10^{19}\,\cms$.
Interestingly, such parameters result in a slab which in all cases is
``marginally'' ionization-bounded with less than 10\% of the ionizing
photons {\it not} absorbed.

The monotonic increase of the \nciv\ column with \up\ is in part due
to the increasing fraction of \civ\ but mostly it is the result of the
slab getting thicker (larger \nhii\ at constant \nhi) since $x_{HI}$
decreases monotonically throughout the slab with increasing \up.  The
slope or curvature of the two solid lines reflect changes in the
internal temperature stratification of the slab with increasing \up.
Because of the dependence of \gamc\ on $T$ (see Eq.~\ref{eqstd}),
there exists  an indirect dependence of \gamc\ on the {\it total}
metallicity given that the equilibrium temperature is governed by
collisional excitation of metal lines (when $Z \gg 0.005$).

The striking result from the slab calculations in Fig.~\ref{omeg} is
that the models with solar metallicity are still two order of
magnitudes below the observed \gamo.  Another way of looking at this
discrepancy is to consider separately the ${I_{CIV}\over
I_{Ly\alpha}}$ emission ratio or the ${N_{HI} \over N_{CIV}}$
column ratio.  Forgetting \gam, just to achieve the observed column of
\nciv\ ($10^{14.5}$ \cms), one would have to use a gas metallicity
below solar by a factor $\ga 50$ (see sequence with $Z=0.02 \zsol$),
which cannot be done without irremediably weakening the \civ\ {\it
emission} line to oblivion. Alternatively, reducing \up\ much below
$10^{-2.5}$ in the solar case can reproduce the \nciv\ column but again the
\civ\ {\it emission} line would be totally negligible.

Might the observed ${I_{CIV}\over I_{Ly\alpha}} = 0.17$ emission line
ratio be anomalous? This is not the case as the observed value in
0943--242 is typical of the value observed in others HZRG without, for
instance, any evidence of dust attenuation of \lya. This ratio is also
that expected from photoionization models if a sufficiently high value
of \up\ is used (Villar-Mart\'\i n et~al. 1996).

Another possibility to consider is the presence of other heating
sources such as shocks which would increase the temperature above the
equilibrium temperature given by photoionization alone.
Alternatively, small condensations in rapid expansion would result in
strong adiabatic cooling and the temperature would be less than given
by cooling from line emission alone. To explore such cases, we have calculated
various isothermal photoionized slabs of different (but uniform)
temperatures (all with $\up = 10^{-2}$). They cover the range
10\,000--40\,000\,K and are represented by the dotted line in
Fig.~\ref{omeg}. These models are in no better agreement with respect
to \gamo. (Varying \up\ for any of these isothermal temperature slabs
would result in an horizontal line). We also computed \gamc\ for a
solar metallicity (precursor) slab submitted to the ionizing flux of a
$500\,\kms$ photoionizing shock (Dopita \& Sutherland 1996). This
model which is represented by an open triangle in Fig.~\ref{omeg} does
not fare better than the power law photoionization models.

\section{Discussion} \label{discussion}

\subsection{Interpretation of the large \gamo\ } \label{intergam}

What is the significance of the obvious discrepancy between models and
the observed \gam? Clearly, the working hypothesis that the emitting
and absorption gas phases are physically the same, is now ruled out
and an alternative explanation must be sought for, based on our result
that the two gas phases (absorption vs. emission) are physically
distinct. We consider the two following explanations in
Sect.~\ref{large} and \ref{twophase}.

\subsubsection{The absorption gas is metal-poor and further out.} \label{large}
Since the absorption gas in this picture is not spatially associated
with the emission gas, its metallicity is unconstrained.  It turns out
that the value of $\gamc\ \simeq 5400$ is easily reproduced by simply
using $\zca/\zce\ \sim 0.005$ in the one-zone case (see
Eqs~\ref{eqemi} and \ref{eqabs}).  The more rigorous stratified slab
geometry would favor a value of $\zca/\zce\ \sim 0.01$ to reproduce
the same \gam, assuming both gas phases to have equal excitation.
Can we disentangle the absolute abundance values? We cannot rely on
the emission spectra alone to derive a precise and independent value
for \zca\ as the emission lines are very model-dependent, with fluxes
from lines like \civ\ depending critically on the temperature. It can
realistically be argued, however, that a
\zce\ less than half solar could {\it not} reproduce the observed metal line
ratios. On the other hand, a \zce\ much higher than solar cannot be
ruled out in absence of direct knowledge of the ionizing continuum
distribution. We consider more plausible a near solar value for \zce\
on the ground that the extended emission lines extend over 13\,kpc and
therefore sample a huge galactic region very distinct from that of the
nucelar BLR (hidden here) which has been shown to be ultra-solar in
high $z$ QSOs (Hamann \& Ferland 1999 and references therein). An
attempt, on the other hand, to model separately the absorption columns
observed in 0943--242 as described below in Sect.~\ref{metal} is more
dependable since temperature is much less of an issue.  The value
inferred below of $\zca \sim 0.01 \zsol$ is consistent with those observed
in absorbers of comparable redshift along the line of sight of more
distant QSOs (Steidel 1990a).  Since measured galactic metallicity gradients
are always negative and a function of the distance to
the nucleus, such a contrast in metallicity between absorption and
emission gas makes more sense if the absorption gas is located much
further out than the emission gas which extends to at least 13\,kpc in
0943--242.

We emphasize that this scenario does not entail that the absorption
gas does not belong to the  environment of the parent
radio galaxy. As argued by vO97, the high frequency of
detection of \hi\ aborbers in 9 out of 10 radio galaxies {\it smaller}
than 50\,kpc, much in excess of the density of absorbers along any
line of sight to distant QSOs, is a compelling argument for concluding
that the absorption  gas is spatially related to the parent galaxy. Our
postulate is that the large scale \hi\ {\it absorption} gas
is the same gas which is seen instead in {\it emission} in those radio
galaxies with \lya\ sizes {\it larger} than 50\,kpc. In effect,
absorption troughs are not seen when the emission gas extends beyond
50\,kpc. Such objects in general also have much larger radio sizes  
as shown by vO97. Kinematically, the gas which is seen in emission at the
largest spatial scales shows narrow FWHM. For instance a reresentative case is
the radio galaxy 1243+036 ($z_e=3.57$) which was studied in great
detail by van Ojik et~al. (1996) and which reveals the presence of
very faint \lya\ emission extending up to 136\,kpc, a region labelled
``outer halo''. This emission gas has a FWHM of 250\,\kms\ and shows
clear evidence for rotational support.

A straightforward explanation of why the same gas is seen in emission
in some objects while in absorption in others might simply be the
environmental pressure.  A larger pressure, like the one adopted by
vO97 can cause the warm gas to condense and hence reduce his filling
factor as compared to similar gas components in a low pressure
environment.  Due to this process, high pressures and consequently
high densities lead to detectable \lya\ since emissivities scale
proportionally to $n_H^2$, but also to an overall smaller covering
factor (hence no detectable absorption) while low pressures lead to
large covering factors (hence absorption) as well as negligible
emissivities. Differences in pressure in the outer halo would therefore
naturally account for the reported dichotomy of detecting
\hi\ troughs exclusively in those emission \lya\ objects devoid of
very large scale emission ($\la 50$\,kpc)

Since absorption troughs tend to be absent in radio galaxies showing
the largest radio scales, we propose that the gas which is seen in
absorption must lie {\it outside} the zone of influence of the radio
jet cocoon, a region with pressure of order
$10^{6}\,\kcmc$ (vO97).  An unpressurized outer halo responsible for
the absorption troughs ought to precede the regime in which the radio
material has expanded sufficiently outward to pressurize the outer
halo.  The eventual increase in environmental pressure would either
disrupt the gas or compresses it into small clumps (making it
unobservable in absorption when the covering factor dwindles),
which  becomes visible in emission if it lies within the ionizing
cone.  vO97 assumed that the absorption and emission gas were
both immersed in zones of comparable surrounding pressure ($n_{\rm H}
T \sim 10^6$ \kcmc) and were therefore of comparable density ($\sim
100$\,\cmc\ for a photoionized gas). We propose instead that whenever
aborption troughs are observed, the absorption gas must lie outside the
radio jet cocoon, allowing for a lower density and high covering
factor.

The clear-cut advantages of locating the \hi\ absorber 
in an unpressurized outer halo are threefold:
\begin{enumerate}

\item We  can now get the high excitation of the low density absorption
gas for free.  In effect, if the density of the absorption gas is as
low as $10^{-3}$--$10^{-2}$\,\cmc, the metagalactic background
radiation suffices to photoionize the absorption gas to the high
degree observed in 0943--242, whether it does or does not lie within
the ionizing cone of the nucleus.  Conversely, for the objects devoid
of absorption, when a higher pressure has set in in the outer halo (as
we presume to be the case in 1243+036), the gas is much denser and can
be seen in emission only if it lies whithin the ionizing cone (since a
high density gas of $\sim 100$\,\cmc\ cannot be kept highly ionized by
the background metagalactic radiation).  This picture would be in
accord with the findings of van Ojik et~al.  (1996) who detect \lya\
in emission in 1243+036 only along the radio axis (presumably the same
axis as that of the ionizing radiation cone) and {\it not} in the
direction perpendicular to it.

\item The much smaller velocity dispersion  ($b \simeq 45\,\kms$) of the
absorption gas as compared to the emission gas (FWHM$/2.35 \simeq
600\,\kms$, cf. Table~\ref{ewciv}) is more readily explained if the
absorption gas lies undisturbed at relatively large distances from the
parent galaxy.

\item It explains why the absorption (yet ionized) 
gas in 0943--242 is not seen in emission while being more massive than
the inner emission \lya\ gas observed within 13\,kpc.  In effect, the
mass of ionized gas either in emission or absorption around 0943--242
inferred by vO97 are $1.4 \, 10^8 \, \msol$ and $ 10^7 ({ x_{HI} /
y_{HII} })^{-1}$ \msol, respectively.  Adopting the conservative value
of $\left<{x_{HI} / y_{HII}}\right> \simeq 0.03$ (cf.  panel~b in
Fig.~\ref{omeg}), the total ionized mass of the absorption ionized gas
therefore exceed that of the inner emission gas by at least a factor
two and yet it is not seen in emission!  This huge pool of ionized gas
can remain undetectable in emission only if it has a very low
density, as argued above.  It is customary to assume a volume filling
factor of $10^{-5}$ for the gas detected in emission in radio galaxies
and that this gas is immersed in a region characterized by a pressure
of order $10^{6}\,\kcmc$ (vO97; van Ojik 1996).  If we suppose
instances where the outer halo has much lower pressure than this, it
can be shown that for the same outer halo mass, the luminosity in
\lya\ would scale inversely to the volume filling factor.  Hence, the gas
would be weaker in emission by a factor of $10^{-5}$ if its filling
factor approached unity (with the mean density being lower by the same
amount).  This scheme would easily explain why the outer halo of
0943--242 is not seen in emission despite its huge mass (comparable
incidentally to the outer halo mass measured in emission in 1243+036 of
$2.8 \, 10^8 \, \msol$ by Ojik et~al.  1996).

\end{enumerate}

\subsubsection{A two-phase gas medium} \label{twophase}
Due to radiative cooling (which goes as $n_H^2$ and rise steeply with
$T$), density enhancements can condense out of the emitting gas and
form a population of about 100 times denser and 100 times cooler
clouds in pressure equilibrium with the ambient medium.  If we
maintain that the pressure characterizing the absorption and the
emission gas is comparable ($\sim 10^6\,\kcmc$) and that either gas
phase has a temperature typical of photoionization, $T\sim 10^4 \,$K,
we obtain (adopting a similar notation to vO97 but adapted to the case
of 0943--242) that the size and the number of small {\em homogeneous}
absorbing condensations required to cover the emission region would be
$0.85 r_{03} \,$pc and $2.4 \, 10^8 r_{03}^{-2} $ clouds,
respectively, where $r_{03} = 0.03~~~$ $\times  \left<{x_{HI} /
y_{HII}}\right>^{-1}$ [as above we adopt 0.03 as the reference neutral H
fraction].  Can we find an alternative interpretation to (1) above for
explaining the large \gamo\ that does not require low metallicities
for the absorption gas?  Such a possibility would arise if the \nhi\
column was not directly related to the \nciv\ column. For instance, in
the auto-gravitating absorber model of Petitjean et~al. (1992), which
consists of a self-gravitating gas condensation with a dense neutral
core surrounded by photoionized outer layers, could in principle give
ratios between columns of \hi\ and \civ\ which do not reflect the
abundance ratio but represents rather the average impact parameter for
our line of sight. Of course, these models have to be rescaled to a
pressure of $n_{\rm H} T = 10^6\,\kcmc$ implying much smaller sizes
but requiring much higher ionizing fluxes (both by a factor $\sim 10^4$).  
This rescaling poses no conceptual problems if we
assume that the photoionization is by the central AGN.  
Using their Figures and Table~4 (Petitjean
et~al. 1992), we infer that the number of auto-gravitating
condensations needed to achieve a covering factor of unity and a mean
\hi\ column of $10^{19}\,\cms$ would have to be large, in excess
of $10^{9.5}$, for instance, for the model C$^{10}_{7000}$. However,
after inspection of the various \nciv\ columns derived from their
extensive grid of models, we did not find any model which would
reproduce the observed \civ\ column without having a metallicity $\le
0.1 \zsol$. The gain in $Z$ is therefore insufficient to get $\zca\
\simeq \zce\ > 0.5$ and we conclude that this explanation for a high
\gamo\ is unworkable.

\subsection{Metallicity determination of the absorption gas} \label{metal}

Our favoured interpretation of the large \gamo\ is that the absorption
gas is of very low metallicity compared to the (inner/denser) emission
gas.  Furthermore, a close parallel in the physical conditions of the
absorption gas could be made with those adopted for the study of QSO
absorbers (e.g.  Steidel 1990a,b; Bergeron \& Stasi\'nska 1986),
namely the densities, the metallicities and the excitation mechanism
(photoionization by a hard metagalactic background radiation).  The
observed \nhi\ column of $10^{19}\,\cms$ would position the 0943--242
absorber in the category of ``Lyman limit system'' according to
Steidel (1992).  The coincidence in physical conditions might be
fortuitous and it does not imply per\,se a common origin or
correspondance between QSO absorbers and outer halos of radio
galaxies.  Under the sole assumption of similar physical conditions,
what estimate of the metallicity can we derive for C?  From the
\nciv/\nhi\ ratio, we cannot determine the ionization parameter and
therefore directly apply the results and models of Steidel (1990b) who
determined for each Lyman limit system a probable range of \up\ from
upper limits or from measurements of other species than \civ.  It is
nevertheless reasonable to assume that the excitation degree in
0943--242 is comparable to that encountered in high excitation QSO
absorbers.  To determine an appropriate value for \up, we adopted the
set of data provided by the three Lyman limit systems observed in the
spectrum of the QSO HS1700+6416 by Vogel \& Reimers (1993) who
successfully measured the columns of up to 3--4 ionization species of
each of the three elements C, N and O.  Amongst our $\alpha =-1 $
model sequence (Sect.\,\ref{upseq}), we selected the model which had
the same \up\ ($\simeq 0.007$) as Vogel \& Reimers (1993) and inferred
that the observed columns in 0943--242 implied that the Carbon
metallicity of the absorption gas was 1\% 
solar (that is C/H $\sim 4 \, 10^{-6}$), which is broadly 
consistent with the range of \zca\ values favored in
Sect.\,\ref{large}.

\subsection{Mean density and cloud sizes}

What would be the minimum density assuming the absorption gas to be
uniformly distributed? If our proposed picture was correct, a
representative size for the absorption gas volume is that given by the
outer halo as seen in emission in other HZRG. Let us adopt the value
measured for 1243+036 by van~Ojik et~al. (1996) of 136\,kpc. Assuming
the same mean ionization parameter as used above (0.007), we derive a
total gas column  of $\nh = \nhii \simeq 10^{21}\,\cms$. Hence
the mean density for a volume filling factor unity on a  scale of the
1243+036 outer halo would be $\simeq 2 \, 10^{-3} \,\cmc$ which is a value
sufficiently low to allow photoionization by the feeble ionizing
metagalactic background radiation.

\subsection{Comparison with the metallicity of BAL QSOs}

Our estimate of the metallicity for the outer halo of 0943--242 is at
odds with the super-solar metallicities (e.g. Hamann 1997, Papovich
et~al. 2000) of the ``associated'' absorbers seen in high redshift
QSOs. The QSO emission gas itself (the BLR) is similarly characterized
by super-solar metallicities (cf. Hamann \& Ferland 1999 and
references therein).  If we consider QSOs and HZRG as equivalent
phenomena observed at different angles, it may  appear at first
surprizing that the metallicities of the absorption components are so
different.  However, we show below that this contradiction is only
apparent as we are probably dealing with totally different gas
components.

\begin{enumerate}
\item {\it Kinematics.} The HZRG large scale absorbers  
are kinematically very quiescent. In effect, the modulus of the
velocity offset between the absorbers and the parent galaxy is usually
less than 400\,\kms\ for the dominant absorber (vO97)\footnote{Highly
blueshifted P-cygni profiles are now known to exist in
radio galaxies with $z \ge 3.5$ (Dey 1999).}.  A substantial
fraction of HZRG absorbers are actually infalling (Binette
et~al. 1998).  This is far from being the case for QSO ``associated''
absorbers whose ejection velocities can extend up to many thousands
\kms\ (Hamann \& Ferland 1999). For instance, the two associated
systems (with detected metal lines) recently studied by Papovich
et~al. (2000) are blueshifted by 680 and 4900\,\kms, respectively.

\item {\it Selection effect.} 
QSOs are spatially unresolved with a size of the source light beam
less than a few light-weeks across. In the case of HZRG absorbers, the
backgound source is the emission gas which extends over a scale $\sim
35$\,kpc. This huge difference in scale results in a totally different
bias on what is preferentially observed. In effect, the extended
absorbers of HZRG are weighted towards the largest volumes and hence
towards the most massive gas components (the total mass of the
absorption component exceeds $10^8 \, \msol$ in 0943--242). By
contrast, in the case of QSO associated absorbers, the mass of gas
directly seen in absorption is tiny (e.g. $\sim 4 \, 10^{-6} \,\msol$
if one considers a background light beam one light-month diameter and
a total gas absorption column of $10^{18}\,\cms$).

\item {\it Coexistence with the BLR.} 
To the extent that QSO associated absorbers represent gas components
expelled from the BLR, we should not be surprized that their
metallicity turn out comparable to the BLR. Given that in HZRG we do
not directly see the pointlike AGN, we cannot expect to see any BLR
component in absorption. As for the extended gas detected in HZRG,
there exists no evidence in favour of super-solar metallicities on
large scales $> 10$\,kpc\ (\nv\ when detected is strong only in
the nucleus)
If a fraction of associated absorbers correspond  to
intervening galaxies close to the QSO, we might expect to see amongst
counterpart HZRGs one or more \civ\ or
\lya\ absorbers of small spatial extent relative to the size of the extended
emission gas.  The weak \hi\ absorption found by Chambers
et~al. (1990) in 4C41.17 might be such occurrence given its partial
coverage of the \lya\ background.
\end{enumerate}

We conclude that HZRG absorbers, when their size is comparable to
galactic halos (as those found by vO97), have probably little to do
with QSO associated absorbers. A more suitable analogy to the
absorption gas of HZRG is that of the Francis cluster of galaxies at
$z=2.38$ which is characterized by large scale absorption gas on a
scale of $\ga 4$\,Mpc (Francis et~al. 2000).

\subsection{Constraints on radio galaxy evolution}

The size of the radio source can be used as a clock that measures the
time elapsed since the start of the radio activity.  A number of
observed characteristics of distant radio galaxies change as a
function of radio size, -- ie. as function of time elapsed
(cf. R\"ottgering et~al. 2000).  For $z\sim 1$ 3CR radio
sources, these include  optical morphology (Best et~al. 1996),
degree of ionisation, velocity dispersion and gas kinematics (Best
et~al. 2000). At higher redshifts ($z>2$), only the smaller radio
galaxies are affected by \hi\ absorption  (vO97). 
All these observations seem to dictate an evolutionary scenario in
which the radio jet has a dramatic impact on its environment while
advancing on its way out of the host galaxy (Rottgering et~al. 2000,
Best et~al. 2000).  

\section{Conclusions} \label{conclusions}

The detection of \civ\ absorption in radio galaxy 0943--242 at the
same redshift as the deep \lya\ trough observed by RO95 demonstrates
that the detected absorption gas is highly ionized.  Having assumed
that the \hi\ and \civ\ columns measured from the Voigt profile
fitting were representative of the dominant gas phase (by mass) in the
outer halo, we have effectively ruled out that the absorption and
emission gas occupy the same position in 0943--242.  We subsequently
reassessed the picture proposed by vO97 in which both the large scale
emission gas and the absorption gas were of comparable density ($n_H
\sim 100\,\cmc$).  In the former picture, the absorption gas was
believed to lie outside the AGN ionization bicone (see their Fig.~11
in vO97). To ionize the gas to such a degree without using the AGN
flux is problematic.  We have proposed an alternative picture in
which the absorption gas is of very low metallicity and lies far away
(in the outer halo) from the inner pressurized radio jet cocoon.
Since in this new scheme the density of the absorption gas is expected
to be very low, the metagalactic background radiation now suffices to
photoionize it.  Furthermore, the structure of the absorption gas is
now drastically simplified since we do not need over $\sim 10^{10}$
condensations of size $\sim 1$pc and density $\sim 10^2\,\cmc$ to
reach a covering factor close to unity. We can now reach similarly
high covering factor using a single or few shells of very low density
which have a volume filling factor close to unity (assuming a density
of $\sim 10^{-2.5}\,\cmc$).

It appears to us that the low metallicity inferred ($Z \simeq 0.01
\zsol$) and the proposed location of the absorption gas in 0943--242
--outside the radio cocoon, in an outer halo which is  seen in
emission in other radio galaxies (as in 1243+036)-- strongly suggest
that the absorbers' existence precedes the observed AGN phase.  Unless
this non-primordial gas has been enriched by still undetected pop III
stars, we consider that it more likely corresponds to a vestige gas
phase expelled from the parent galaxy during the initial starburst at
the onset of its formation.

If the \civ\ doublet was detected in absorption in other radio
galaxies with deep \lya\ absorption troughs, there are many aspects
which would be worth studying.  For instance, how uniform is the
excitation of the absorption gas across the region over which it is
detected?  Is a single phase sufficient? This could be tested by an
attempt to detect absorption troughs of \mgiiw\ or imaging the troughs
in \civ\ with an integral field spectrograph on an 8-m class
telescope.  How different is the metallicity of the absorption gas in
the other radio galaxies?  The information gathered could then be used
to infer the enrichment history of the outer halo gas which surrounds
HZRG.

\begin{acknowledgements}
We are grateful for the referee's comments which raised many
interesting issues we had overlooked.  We thank Richard Hunstead and
Joanne Baker for taking part in the observations.  One of the authors
(LB) acknowledges financial support from  CONACyT grant 27546-E.

\end{acknowledgements}

\end{document}